\documentclass{elsart}
\usepackage{amsmath,amssymb,epsfig}

\begin{document}

\begin{frontmatter}

\title{Pair Formation within Multi-Agent Populations\thanksref{label1}}
 \thanks[label1]{We thank EPSRC (U.K.) for funding though grant EP/D001382, and the  
European Union for funding under the MMCOMNET programme.}

 \author{David M.D. Smith}, 
\ead{david.smith@lincoln.ox.ac.uk}
\author{Neil F. Johnson}
\ead{n.johnson@physics.ox.ac.uk}

\address{Physics Department, Clarendon Laboratory, Parks Road, Oxford, OX1 3PU, U.K.}

\begin{abstract}
We present a simple model for the formation of pairs in multi-agent populations of type $\bf A$ and $\bf B$ which move freely on a spatial network. Each agent of population {\bf A} (and {\bf B}) is labeled as $A_i$ (and $B_j$) with $i=1, \dots, N_A$ (and $j=1, \dots, N_B$) and carries its own individual list of characteristics or `phenotype'. When agents from opposite populations encounter one another on the network, they can form a relationship if not already engaged in one. The length of time for which any given pair stays together depends on the compatibility of the two constituent agents. Possible applications include the human dating scenario, and the commercial domain where two types of businesses $\bf A$ and $\bf B$ have members of each type looking for a business partner, i.e. $A_i+B_j\rightarrow R_{ij}$. The pair $R_{ij}$ then survives for some finite time before dissociating $R_{ij}\rightarrow A_i+B_j$. There are many possible generalizations of this basic setup. Here we content ourselves with some initial numerical results for the simplest of network topologies, together with some accompanying analytic analysis.
\end{abstract}

\end{frontmatter}

\section{Introduction}
The formation of pairs or `relationships' between members of two distinct populations, is a phenomenon which is of central importance in a number of application domains: from animal and human societies through to commerce and politics. Here we consider a simple yet highly non-trivial model of such a dynamical process, in which we allow such pairs to form and break up based on an appropriate measure of their mutual compatibility. Although there are many possible variations and generalizations according to the particular real-world system of interest, we content ourselves here with an analysis of the basic model itself, both numerically and analytically.

Specifically, we consider a system comprising two populations 
$\bf A$ and $\bf B$. Population $\bf A$ contains $N_A$ agents, while population $\bf B$ contains $N_B$ agents. All agents are assumed to be able to move freely on a network. In future work, we will explore the effects of specific network topologies which are designed  to mimic the real-world social/business space within which such agents move, for example scale-free or small-world networks \cite{Dorogovtsev}. However in the present paper, we focus on results which are either generic to all networks or which hold for a specific simple network. 

As the agents move on the network, they will interact with each other on a given timestep if they happen to find themselves on the same node. Members of each population wish to form a pair or `relationship' with a member of the other population. We allow the members of each population to have their own list of characteristics (or equivalently, preference list or `phenotype'). Upon finding each other, a given pair will stay together a finite amount of time which depends on the similarity of their respective lists -- this mimics a relationship whose duration depends on the relative compatibility of the individual agents comprising the pair. In other words, $A_i+B_j\rightarrow R_{ij}$ where the pair $R_{ij}$ survives for some finite time before dissociating $R_{ij}\rightarrow A_i+B_j$.
The goal of each individual is to maximize the time it spends in a partnership. One interesting  application is in the human dating scenario comprising the two distinct populations (male and female) where members of each sex are looking for a partner with which to form a relationship. 

The layout of the paper is as follows: Section 2 gives further details of the model. Section 3 provides the main results while Section 4 presents the conclusions. We find numerically that there is a highly non-linear relationship between expected utility, the threshold for formation of a relationship, and the degree of sophistication of the individual agents. In order to explain this finding, we develop an analytic theory which depends on the average period of time which an agent spends in a pair and the probability of finding a suitable partner on the network. The agreement between the analytic results and the numerical calculations is good, despite the fact that this problem is more complicated than standard reaction-diffusion problems -- in particular, our particles (i.e. agents) have non-trivial internal degrees of freedom deriving from their preference lists.
Ae mentioned above, we focus here on a basic version of the model although the analysis should be generalizable to any single component network within which the agents explore.

Before proceeding, we comment on the difference between the present model and the so-called `stable marriage problem' \cite{beauty,Sex}. The latter has been investigated many times in the literature, typically using a Nash Equilibrium approach to optimize the local utilities of two lists of agents which are looking to pair up. When this equilibrium exists, any change of the state of the system caused by an agent altering its choice, results in a worse performance for that agent \cite{beauty}. The agents try to minimize their `energy' by marrying someone who fills as many of their criteria as possible. However unlike most traditional stable marriage analysis, we will not look at minimizing the system's `energy' \cite{Sex} as a whole -- our rationale being that in many real-world systems, there is no opportunity for exchange of global information in order to optimize the system. Instead, agents are restricted to interacting with others who occupy the same node thereby accessing only a subset (albeit dynamic) of the whole social network.  Likewise the timescales required to explore and subsequently optimize one's own local utility, are not typically achievable in systems where agents only exist for a finite time.  Moreover in contrast to recent models of business formation and/or collapse \cite{Axtell}, the local utility of an agent in our model who is choosing whether or not to enter a partnership, is not based on the intrinsic value of being instantaneously single -- instead it is based on the expectation of future satisfaction.

\section{Model of Pair Dynamics}
We consider the agents to be moving freely on a spatial network, with each individual placed at a random initial location. The agents undergo a random-walk around this network, occupying a particular site at a given timestep. If there exist one or more single agents from each population on a given node at a given timestep, then one or more relationships might form. In particular, each (single) agent is allowed to interact with one (single) agent chosen randomly from the members of the opposite population who are currently on that same site. There is no sense of optimization in this pair-formation process: as in many real-world scenarios, the agents do not know prior to this interaction how well-suited a potential partner might be. A pair of agents can only find out their `compatibility' after they have interacted, and this is restricted to once per timestep. Hence if, for example, there exist more single agents from population $\bf A$ than from population $\bf B$ on any particular node at a given timestep, then the number of agents which do not interact at all at that timestep is given by this excess number of population $\bf A$ agents.

\begin{figure}[h]
\begin{center}
\includegraphics[width=0.7\textwidth]{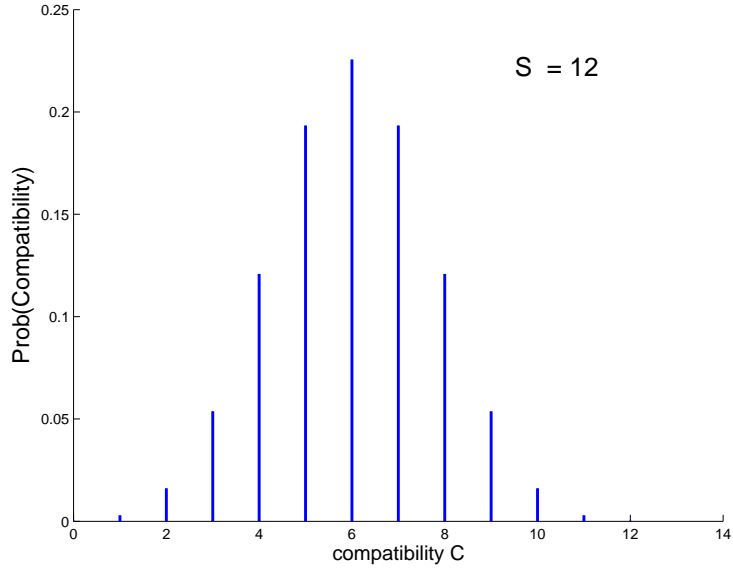}
\end{center}
\caption{The compatibility $C_{ij}$ of two agents which are randomly chosen from populations $\bf A$ and $\bf B$ respectively, where $\bf A$ and $\bf B$ have uniform `phenotype' distributions (i.e. the distribution of the randomly allocated preference lists is uniform). For illustration, we have used $S=12$.}
\label{fig:compat}
\vskip0.2in
\end{figure}

We express the compatibility of two agents $A_i$ and $B_j$ as follows\footnote{This definition of compatibility between agents' preference lists is similar in construction to the similarity of strategies in Binary Agent Resource games such as the Minority Game, where strategies are represented by bit-strings of a particular length. In particular, the similarity of strategies $i$ and $j$ is given by the Hamming distance \cite{Hamming} between their respective bit-strings \cite{Johnson}.}:
\begin{equation}
C_{ij}~ = ~\frac{{\bf a}_i\cdot {\bf b}_j~+~S}{2}
\end{equation}
where ${\bf a}_i$ and ${\bf b}_j$ are vectors representing the preference lists of agents $A_i$ and $B_j$ respectively, with elements restricted to $\pm1$, and $S$ is the length of these lists\footnote{This setup could be generalized by assigning each agent a separate set of preferences {\em and} attributes. The value assigned to the compatibility of a given pair of agents, could then be different for each agent in that pair. The subsequent pair-formation rules would need to be modified to include this feature. We also note that the notion of beauty could be introduced into the model setup, by biasing how the preference lists are allocated at the start of the game.}. The compatibility $C_{ij}$ lies in the region $0\leq C_{ij}\leq S$ and is equal to $S$ minus the Hamming distance \cite{Hamming} between the two vectors. We use the compatibility $C_{ij}$ to prescribe the duration of a relationship between a given pair, in the case that the pair actually forms. Again this could be generalized to allow for exogenous perturbations, e.g. subsequently meeting more compatible partners -- but here we shall assume that all relationships last for the assigned duration. In particular, we here assume that the lifetime of a given pair is a monotonically increasing function of the compatibility $C_{ij}$. In principle, such a lifetime could be generated by making the relationship a dynamic system whereby each agent calculates a utility associated with the relationship, such as the Cobb-Douglas \cite{Axtell} utility function -- and an agent then chooses to leave the pair when the descent to some equilibrium falls below some critical threshold. In the situation whereby a particular single agent is exposed to a random single agent from the other population, the compatibility will be binomially distributed, $C~\epsilon~Bin(S,\frac{1}{2})$ as shown in Fig. \ref{fig:compat}. 

In order to mimic real-world scenarios, we choose a game set-up whereby agents do not enter a relationship if $C_{ij}$ falls below some critical level which we call the compatibility threshold $\tau$. The value $\tau$ could be allowed to vary between agents and/or evolve due to past experiences, with the minimum $\tau$ being the critical one in a given potential pair -- however in this work we focus on the case where all agents have the same $\tau$ and where $\tau$ is time-independent. The goal of each agent is to maximize its own utility, i.e. the time it spends in a relationship. The value associated with being in a relationship in a given timestep is binary (see Fig. 2) with the compatibility then deciding the duration of the relationship.

\begin{figure}[h]
\vskip0.2in
\begin{center}
\includegraphics[width=0.7\textwidth]{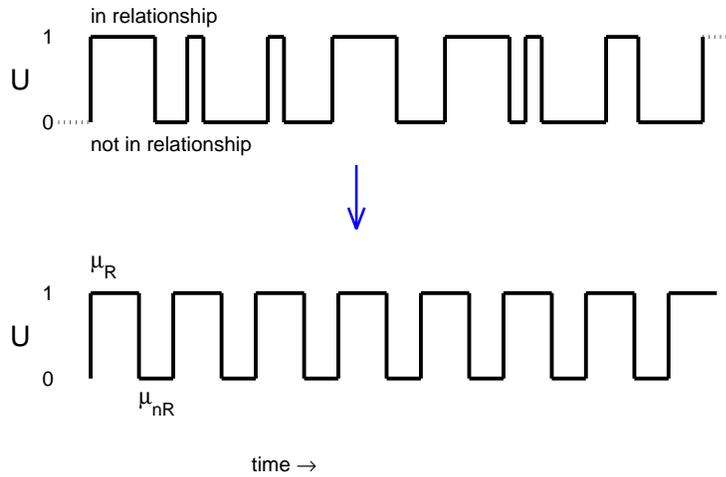}
\end{center}
\caption{The (binary) utility associated with an agent being in a relationship.}
\label{fig:reltime}
\vskip0.2in
\end{figure}

\section{Numerical and Analytic Analysis of Pair Dynamics}

Each agent at any given timestep can either be in a relationship ($R$) or not in a relationship ($nR$). The expected utility per unit time for $\it any$ given agent can be expressed as the average time spent in a relationship over a given time period, as shown schematically in Fig. \ref{fig:reltime}. This can be written as follows:
\begin{equation}
\overline{U} ~=~ \frac{\mu_R}{\mu_R~+~\mu_{nR}}
\label{eqn:simplemain}
\end{equation}
where $\mu_R$ is the expected duration of a relationship and 
$\mu_{nR}$ is the expected duration of any non-relationship spell -- in other words, the expected time before successfully forging a relationship having started in the $nR$ state. This result is general and applies to any agent in an arbitrary network. Assuming random interactions between agents as they wander randomly through the network, the $\mu_R$ term can be calculated from the distribution of $C$ and the function 
$D(C)$, which prescribes the relationship's duration for a given compatibility. For an $nR$ agent, the probability that it will  remain in state $nR$ for a further timestep will be one minus the probability of successfully forging a relationship. As such, the value $\mu_{nR}$ is governed by a geometric probability distribution and can be calculated accordingly:

\begin{equation}
\mu_{nR}~=~ \sum_{\lambda =0}^\infty P_{success}(1-P_{success})^\lambda ~\lambda
\label{eqn:geomet}
\end{equation}

Here $P_{success}$ is the probability of an $nR$ agent from 
$\bf A$ or $\bf B$, interacting with an $nR$ agent from $\bf B$ or $\bf A$ $\it and$ successfully forming a relationship. We can rewrite $P_{success}$ as the probability of interacting with a potential partner $P_{int}$ and the probability of successfully forming a relationship once a potential partner is engaged (i.e. $P_R$). Assuming random interactions, the latter is merely a summation based upon the compatibilities which would allow a relationship to happen (given the threshold $\tau$) and their respective likelihoods. As such, we can write  $\mu_{R}$ and 
$\mu_{nR}$ in terms of  $P_{int}$ and $P_R$:
\begin{eqnarray}
\mu_{nR}&=&\sum_{\lambda =0}^\infty P_{int} P_R(1-P_{int}P_R)^\lambda~\lambda \nonumber\\
{}&=& \frac{1- P_{int}P_R}{P_{int}P_R}    \label{eqn:geosum}\\
{\rm and}&{}&{}\nonumber\\
\mu_R&=&\frac{\sum_{C =\tau}^S~P(C)~D(C)}{P_R}\label{eqn:muR}\\
{\rm where}&{}&{}\nonumber \\
P_R&=& \sum_{C =\tau}^S~P(C)  \label{eqn:PR}\ \ .
\end{eqnarray}

\begin{figure}[h]
\begin{center}
\includegraphics[width=0.7\textwidth]{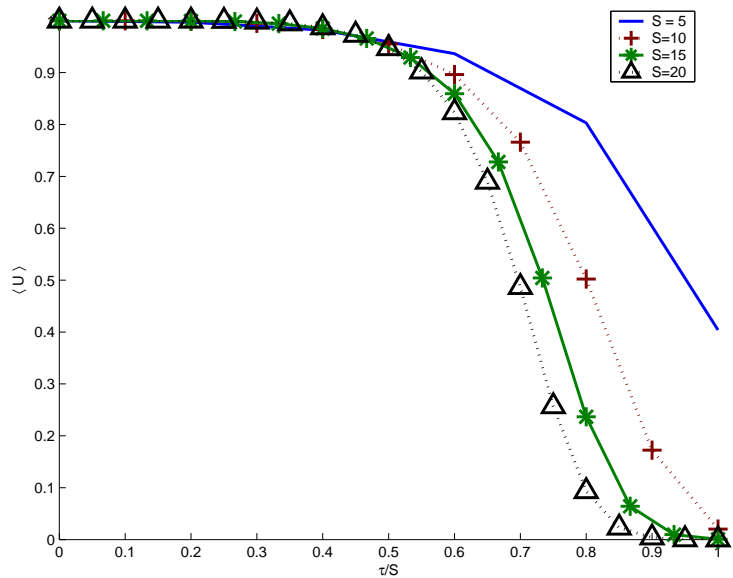}
\end{center}
\caption{The average utility $\langle U \rangle$ as a function of 
$\tau/S$. This shows the importance of $\tau$ with increasingly sophisticated populations, i.e. longer attribute lists and hence larger $S$. These curves are calculated for a network comprising just a single node, hence the interaction probability $P_{int} = 1$. Also $\alpha = 20$ and $\beta~=~1$.}
\label{fig:Usurf2}
\vskip0.2in
\end{figure}

We note that the effect of the movements of the agents on any particular network, is embodied in the probability term  $P_{int}$. As an example, we consider the simplest topology of network -- one node - and a simple function for the pair lifetime $D(C)~=~\alpha \big(\frac{C}{S}\big)~+~\beta$. All agents' expected interactions are equal for this system, hence the mean utility over all agents will be the same and is  given by $\langle{U}\rangle_{agents}~=~\overline{U}$. With $N_A~=~N_B~ =~N $, the average time required for meeting a potential partner by an agent who has been in state $nR$ (because of failing to overcome the threshold) is unity since $N_A=N_B$ and all agents are in the same place. As such, $P_{int}=1$. Substituting Eq. \ref{eqn:geosum} and Eq. \ref{eqn:muR} into Eq. \ref{eqn:simplemain}, the expected utility per timestep averaged over all agents can be written:
\begin{eqnarray}
\langle U \rangle &=& \frac{P_R\mu_R}{P_R\mu_R~+~1-P_R} \nonumber\\
P_R&=& \sum_{C =\tau}^S~ \left(\begin{array}{c}
S\\ 
C 
\end{array}\right) \left(\frac{1}{2}\right)^S  \ \ .
\end{eqnarray}
The dependence on the threshold $\tau$ will become more important for more sophisticated agents (i.e. higher $S$  and hence longer attribute lists) since the likelihood of meeting a perfectly compatible agent then becomes low. This effect can been seen in Fig. \ref{fig:Usurf2} which indicates the highly non-linear relationship between the average utility, the threshold $\tau$ and the level of sophistication $S$.

For a more general network structure, we must calculate or infer from numerical simulations the interaction probability of those agents not in a relationship, i.e. $P_{int}$. This term will depend on the details of how the agents move and the topology of the particular network. It will also be a function of the number of potential partners available, and hence is linked to the initial number of each population ($N_A$, $N_B$) and the fraction of these who are currently in a pair. If the agents are moving randomly on a highly-connected or crowded network, then the interactions between particular phenotypes will be random. This will also be true for a situation in which agents forming a pair do not physically stick to each other,  but rather roam independently throughout the lifetime of that relationship. Under these conditions, $P(C)$ will again be binomially distributed and the utility of the individual over time can be calculated. Again, all agents' expected interactions are equal so $\langle{U}\rangle_{agents}~=~\overline{U}$. 

We now consider the situation in which the time-averaged number of single agents from each population on any node is uniform over the entire network. This would be appropriate to networks where the degree of each node is identical (e.g. the fully connected graph). Future work will discuss relaxing this restriction to general topologies.
Specifically, we focus on two regimes: overcrowded and undercrowded. In the overcrowded situation, there are more potential partners than nodes. As each agent can only test for compatibility once per timestep, any difference in the number of agents from each population on a particular node at a particular time will result in a group which cannot interact with a potential partner. For example, if there are 5 type-$\bf A$ agents and 3 type-$\bf B$ agents in state $nR$ on a particular site, then 2 type-$\bf A$ agents will not be able to interact with a potential partner. If there are large fractions of the populations in state $nR$ as compared to the number of nodes, we can approximate the (binomial) distributions of each population on a particular site by Normal distributions. For a total of $N$ agents of each population, the number in state $nR$ will be $N(1-\langle U \rangle)$. By integrating over the distribution of the difference of $nR$ agents from each population on a particular node, we can calculate the subsequent expectation number of non-interacting agents. On a network of $n$ nodes, we can then write the probability of interaction for a given agent of either population as follows \cite{website}: 
\begin{eqnarray}
P_{int}&=&1-\left(\frac{n-1}{\pi N(1-\langle U\rangle)}\right)^{\frac{1}{2}}
\label{eqn:over}
\end{eqnarray}

\begin{figure}[h]
\begin{center}
\includegraphics[width=1.0\textwidth]{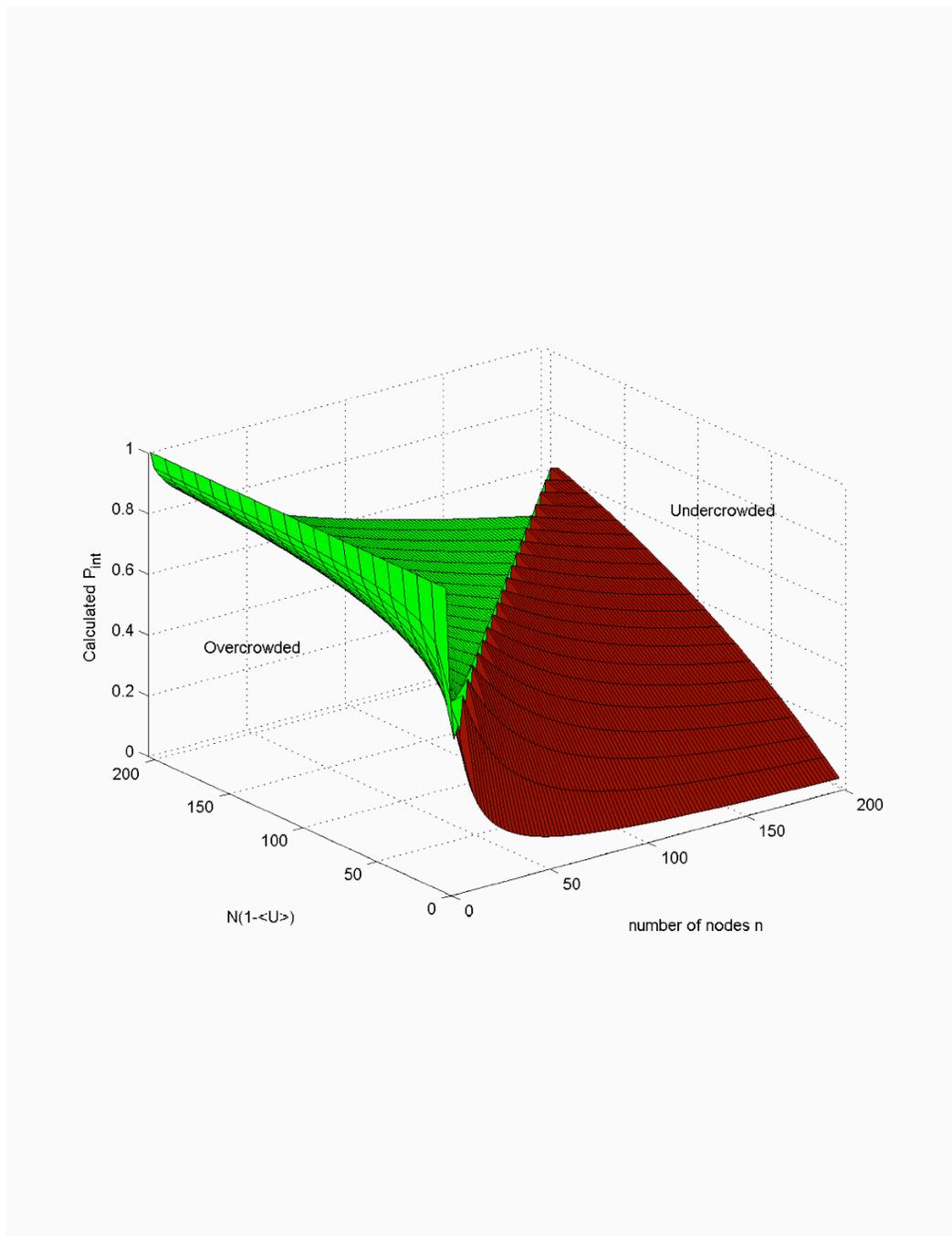}
\end{center}
\caption{Analytic results for the interaction probability $P_{int}$ for the overcrowded and undercrowded regimes.}
\label{fig:calcprob}
\end{figure}

In the undercrowded regime, the probability of having more than one agent of a given population on a particular node is small. Hence the probability of a given agent meeting a potential partner can be described as follows (see Ref. \cite{website} for details):
\begin{eqnarray}
P_{int}&=&1-\left(\frac{n-1}{n}\right)^{N(1-\langle U\rangle)}
\label{eqn:under}
\end{eqnarray}
These two approximations are only strictly valid in the limits $N(1-\langle U\rangle)\gg n$ and $N(1-\langle U\rangle)\ll n$ respectively. However, in what follows we will show that these expressions turn out to be reliable in practice over the wider ranges $N(1-\langle U\rangle)\geq n$ and $N(1-\langle U\rangle)\leq n$. Figure \ref{fig:calcprob} shows the values of  $P_{int}$ for the two ranges, calculated using the two respective expressions.
Substituting Eq. \ref{eqn:geosum} into Eq. \ref{eqn:simplemain} yields the expected utility for agents in this type of network:
\begin{equation}
\langle U \rangle ~=~ \frac{\mu_R}{\mu_R~+\frac{1-P_{int}P_R}{P_{int}P_R}}\ \ .
\label{eqn:main}
\end{equation}
Comparison of Eq. \ref{eqn:main} to the expressions for $P_{int}$ (Eq. \ref{eqn:over} and Eq. \ref{eqn:under}) results in two sets of simultaneous equations (one set for each regime) which we can then solve numerically for $\langle U \rangle$ for a given network.

\begin{figure}[h]
\begin{center}
\includegraphics[width=1.1\textwidth]{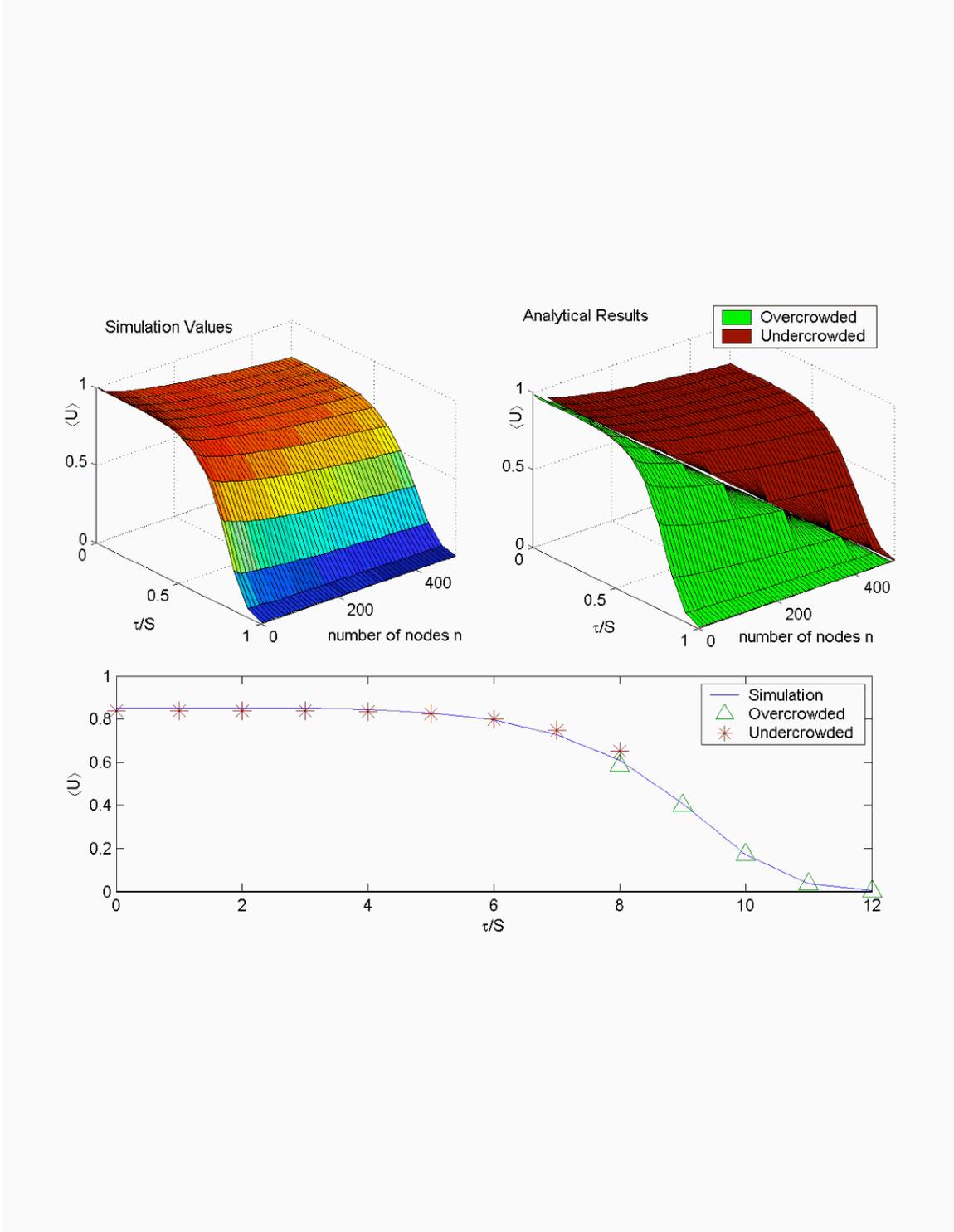}
\end{center}
\caption{Comparison of the values of $\langle U\rangle$ obtained from numerical simulation, to those obtained using the analytic expressions in the overcrowded and undercrowded regimes. The phenotype (i.e. preference list) length is $S = 12$. The total number of agents from each population is $N=500$, while $\alpha=20$ and $\beta~=~1$. The simulation is performed on a fully-connected graph. }
\label{fig:3in1}
\end{figure}

Figure \ref{fig:3in1} shows that the results obtained using the analytic expressions do indeed compare very well to those from numerical simulations. The numerical simulations are performed here on a fully connected network. This positive result gives us confidence that similarly accurate results might be obtained for more general networks. We will explore this in a future publication. 

\newpage
\section{Conclusion}
We have introduced and analyzed a simple model of agent interactions on a network, using a binary utility function to describe the lifetime of a given pair. We have focused on the specific case in which the system is in a mixed state such that the interactions between agents are not biased. The analysis can however be generalized to any network topology and payoff function, by considering the corresponding interaction probabilities and utility weighting. Further provision could be made to include biased interactions (e.g. where an individual is more likely to encounter its previous partner).
If we consider a specific agent who wants to maximize its local utility in such a situation, changing phenotype will have no effect -- however a conscious decision to remain at a particular hub ($P_{int}\rightarrow 1$) could have a dramatic effect.

\end{document}